\RequirePackage{fix-cm}
\documentclass{llncs}       %
\usepackage[table]{xcolor}

\usepackage{graphicx}
\usepackage{xspace}
\usepackage{hyperref}
\usepackage{wrapfig}

\hypersetup{
    colorlinks,
    linkcolor={red!50!black},
    citecolor={blue!50!black},
    urlcolor={blue!80!black}
}

\graphicspath{{./images/}{../images/}}
\newcommand{\dL}{d$\mathcal{L}$\xspace}
\usepackage{tikz}
\usetikzlibrary{arrows,positioning}
\usepackage[super]{nth}
\usepackage{float}
\tikzset{
    >=stealth',
     main/.style={
           rectangle,
           very thick,
           draw=black,
           text width=7em,
           minimum height=2em,
           text centered},
     supp/.style={
           rectangle,
           draw=black,
           text width=6.5em,
           minimum height=2em,
           text centered},
    arrow/.style={
           ->,
           thick,
           shorten <=2pt,
           shorten >=2pt,}
}
\usepackage{subfiles}

\usepackage{syntax}

\usepackage{amsmath}
\usepackage{amssymb}
\usepackage{pifont}
\usepackage{threeparttable}
\usepackage{array}

\usepackage{bussproofs}
\usepackage{alltt}

\makeatletter
\newcommand{\labelthis}[2]{%
  \def\@currentlabel{#2}\label{#1}{\scriptsize \textnormal{(#2)}}%
}
\makeatother

\newcommand{\yes}{\cellcolor[HTML]{67FD9A}\ding{51}}
\newcommand{\no}{\cellcolor[HTML]{FD6864}\ding{55}}

\newcommand{\real}{\mathbb{R}}
\newcommand{\skey}[1]{\textit{\textsf{#1}}}
\newcommand{\defs}{\triangleq}
\newcommand{\hvderiv}[4]{#1 \rhd #2 \,[#3 \in #4]}

\newtheorem{nonformat-algorithm}{Algorithm}%
\usepackage{algorithm2e}

\begin{document}

\title{Certifying Differential Equation Solutions from Computer Algebra Systems in Isabelle/HOL}

\author{Thomas Hickman \and Christian Pardillo Laursen \and Simon Foster
}

\institute{University of York}

\date{Received: date / Accepted: date}

\maketitle

\begin{abstract}
The Isabelle/HOL proof assistant has a powerful library for continuous analysis, which provides the foundation for verification of hybrid systems. However, Isabelle lacks automated proof support for continuous artifacts, which means that verification is often manual. In contrast, Computer Algebra Systems (CAS), such as Mathematica and SageMath, contain a wealth of efficient algorithms for matrices, differential equations, and other related artifacts. Nevertheless, these algorithms are not verified, and thus their outputs cannot, of themselves, be trusted for use in a safety critical system. In this paper we integrate two CAS systems into Isabelle, with the aim of certifying symbolic solutions to ordinary differential equations. This supports a verification technique that is both automated and trustworthy.
\end{abstract}

\section{Introduction}
Verification of Cyber-Physical and Autonomous Systems requires that we can verify both discrete control, and continuous evolution, as envisaged by the hybrid systems domain~\cite{alur2011}. Whilst powerful bespoke verification tools exist, such as the KeYmaera X~\cite{KeYmaera} proof assistant, software engineering requires a general framework, which can support a variety of notations and paradigms~\cite{Gleirscher2018-NewOpportunitiesIntegrated}. Isabelle/HOL~\cite{Isabelle} is a such a framework. Its combination of an extensible frontend for syntax processing, and a plug-in oriented backend, based in ML, which supports a wealth of heterogeneous semantic models and proof tools, supports a flexible platform for software development, verification, and assurance~\cite{Wenzel2007FMIsabelle,Brucker2019-DOFCert,Foster2020AUV}.

Verification of hybrid systems in Isabelle is supported by several detailed libraries of Analysis, including \textsf{Multivariate Analysis}~\cite{Harrison2005-Euclidean}, \textsf{Affine Arithmetic}~\cite{immler2018}, and \textsf{HOL-ODE}~\cite{immler2012}, which supports reasoning for Systems of Ordinary Differential Equations (SODEs). These libraries essentially build all of calculus from the ground up, and thus provide the highest level of rigour. However, Isabelle currently does not offer many automated proof facilities for hybrid systems. KeYmaera X~\cite{KeYmaera}, in contrast, is highly automated and thus very usable, even by non-experts. This is partly due to the inclusion of efficient algorithms for differential equation solving and quantifier elimination, which are both vital techniques. Several of these techniques are supported by integration with Computer Algebra Systems (CAS), which support these and several other algorithms, in particular the Wolfram Engine, which is the foundation for Mathematica.

Nevertheless, whilst CAS systems are efficient, they do not achieve the same level of rigour as Isabelle, and thus the results cannot be used without care in a high assurance development. In Isabelle, all results are certified using a small kernel against the core axioms of the object logic, following the LCF architecture. The correctness of external tools does not need to be demonstrated, but only that particular results can be certified. This approach has been highly successful, and in particular has allowed the creation of the famous \textsf{sledgehammer} tool~\cite{Blanchette2011,Blanchette2016Hammers}, which soundly integrates external automated theorem proving systems.

In this paper, we apply this approach to integration of CAS systems into Isabelle/HOL\footnote{The code supporting our approach can be found in the following GitHub repository: \url{https://github.com/ThomasHickman/Isabelle-CAS-Integration}}. We focus on generation and certification of solutions to SODEs, though our approach is more generally applicable. We integrate two CAS systems: the Wolfram Engine and SageMath, the latter of which is open source. We show how SODEs and their solutions can be described and certified in Isabelle. We then show how we have integrated the two CAS systems, using their APIs, and several new high level Isabelle commands. We evaluate our approach using a large test set of SODEs, including a large fragment of the KeYmaera X example library. Our approach is largely successful, but we highlight some future work for improving the certification proof process in Isabelle.

The structure of our paper is as follows. In \S\ref{sec:background}, we highlight related work and necessary context. In \S\ref{sec:ode-cert}, we describe our tactic for certification of SODEs. In \S\ref{sec:sagemath} and \S\ref{sec:wolfram}, we present our integrations with SageMath and Wolfram, respectively. In \S\ref{sec:evaluation}, we evaluate our approach using our test set, and in \S\ref{sec:concl} we conclude.

\section{Background}
\label{sec:background}

The dominant approach for CPS verification is differential dynamic logic (\dL), a proof calculus for reasoning about hybrid programs~\cite{dL}. Hybrid programs allow modelling of hybrid systems by providing operators for discrete transitions, such as assignment and nondeterministic composition, together with modelling dynamics via continuous evolution of a SODE.

The most advanced tool for deductive verification of hybrid systems is KeYmaera X~\cite{KeYmaera}, a theorem prover for \dL. Its capabilities have been shown in numerous case studies, such as in~\cite{morerobix} for verifying various classes of robotic collision-avoidance algorithms, and in~\cite{acasx} for proving that the ACAS X aircraft collision avoidance system provides safe guidance under a set of assumptions. KeYmaera X uses the Wolfram Engine for SODE solving and quantifier elimination.

KeYmaera X is, however, restricted to reasoning about \dL hybrid programs, and cannot be applied directly to other notations. In particular, we cannot show that a controller specification is refined by a given implementation in a language like C~\cite{Tuong2019-CIsabelle}, although tools such as VeriPhy~\cite{veriphy} and ModelPlex~\cite{modelplex} somewhat bridge this gap. It also cannot handle transcendental functions, such as $\sin$ and $\log$, which are often used by control engineers.

\dL has also been implemented~\cite{verified-dL,Munive2018-DDL,Foster2020-dL} in the Isabelle proof assistant~\cite{Isabelle}, as both a deep~\cite{verified-dL} and shallow embedding~\cite{Munive2018-DDL,Foster2020-dL}. Verification in Isabelle brings the advantage of generality, whereby the hybrid systems proof could be used to show correctness of an implementation~\cite{Tuong2019-CIsabelle}, or used in a larger proof about a complex system. It also allows integration with several notations in a single development, which is the goal of our target verification framework, Isabelle/UTP~\cite{Foster2020-IsabelleUTP}.

A present disadvantage of Isabelle is the lack of automated proof, in comparison to KeYmaera X. Consequently, our goal is to improve automation by safe integration of a CAS. Mathematica has previously been integrated into Isabelle for quantifier elimination problems over univariate polynomials~\cite{Li2017-Poly}. Here, we integrate two CAS systems for the purpose of certifying SODE solutions. 

Plugins in Isabelle, like our CAS integration, are written using the ML language, and manipulate terms of the logic. Terms are used to encode a typed $\lambda$-calculus, and are encoded using the following ML data type.

\begin{alltt}
\textbf{datatype} term = Const \textbf{of} string * typ | Free \textbf{of} string * typ |
    Var \textbf{of} indexname * typ | Bound \textbf{of} int |
    Abs \textbf{of} string * typ * term | \$ \textbf{of} term * term
\end{alltt}
The type \texttt{typ} describes Isabelle types. The basic constructors include constants (\texttt{Const}), free variables (\texttt{Var}), schematic variables (\texttt{Var}), and bound variables (\texttt{Bound}) with de Bruijn indices. With the exception of \texttt{Bound}, these all consist of a name and a type. The remaining two constructors represent $\lambda$-abstractions and applications of one term to another. For example, the term $\lambda x~y : \real. \, x + y$, a function that adds together two real numbers, is represented as follows:
\begin{alltt}
Abs ("x", "real", Abs ("y", "real",
    Const ("Groups.plus_class.plus", "real => real => real") 
      \$ Bound 1 \$ Bound 0))
\end{alltt}
As usual, $\lambda x~y. \, e$ is syntactic sugar for $\lambda x. \lambda y. \, e$. Predefined functions, such as $+$, are represented by constant terms are are fully qualified. 

Our CAS plugin takes as input a SODE encoded as a term, which it turns into input for a CAS. The CAS returns a solution in its own internal representation, if one exists, and this is turned into another Isabelle term, and certification of the solution is attempted. Our approach builds on both on Immler's library for representing SODEs and their solutions~\cite{immler2012,immler2018} (\textsf{HOL-ODE}). 

In the next section we describe the approach for SODE certification. 

\section{Certifying SODE Solutions}
\label{sec:ode-cert}

\begin{figure}[t]
\centering
\includegraphics[width=.8\textwidth]{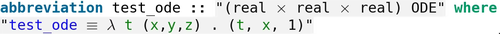}

\vspace{-2ex}

\caption{SODE representation in Isabelle}\label{fig:repr}

\vspace{-3ex}
\end{figure}

In this section we describe how SODE solutions can be certified using our \verb|ode_cert| proof tactic. We assume a SODE of the form $\dot{x}(t) = f~t~(x~t)$, described by a function $f : \real \to \real^n \to \real^n$, which gives a vector of derivatives for each continuous variable at time $t$ and current state $x~t$. A candidate solution to this SODE is a function $x : \real \to \real^n$, which can potentially have a restricted domain $T \subseteq \real$ and range $D \subseteq \real^n$. For example, consider the following SODE:
\begin{example} \label{ex:SODE} $(\dot{x}(t), \dot{y}(t), \dot{z}(t)) = (t, x(t), 1)$
\end{example}
It has three continuous variables, and can be represented with the function $g \defs (\lambda\,t~(x, y, z).\, (t, x, 1)$ whose type is $\real \to \real^3 \to \real^3$. Its representation in Isabelle is shown in Figure~\ref{fig:repr}, where a name is introduced for it by an abbreviation.

The goal of the \texttt{ode\_cert} tactic, then, is to prove conjectures of the form $$x \mathop{\skey{solves-ode}} f~T~D$$ which specifies that $x$ is indeed a solution to $f$, and is defined within the \textsf{HOL-ODE} package~\cite{immler2012}. It requires that we solve the following two predicates:
$$(\forall t \in T.\, (x \mathop{\skey{has-vector-derivative}} ~(f~t~(x~t))~ (\skey{at}~t~\skey{within}~T))) ~\text{and}~ x \in T \to D$$
We need to show that at every $t$ in the domain, the derivative of $x$ matches the one predicted by $f$, and that $x$ has the correct domain and range. The predicate $\skey{has-vector-derivative}$ is defined within the \textsf{HOL-Analysis} package~\cite{Harrison2005-Euclidean}, which also provides a large library of differentiation theorems. For brevity, we use the syntax $\hvderiv{f}{f'}{t}{T}$ to mean $(f \mathop{\skey{has-vector-derivative}} f') (\skey{at}~t~\skey{within}~T)$.

\begin{theorem}[Derivative Introduction Theorems] \label{thm:der-intro} $ $%

\vspace{2ex}

\centering
\begin{tabular}{cccc}

\AxiomC{---}
\RightLabel{\labelthis{rule:a}{a}}
\UnaryInfC{$\hvderiv{(\lambda x.\, c)}{0}{t}{T}$}
\DisplayProof
~~
\AxiomC{---}
\RightLabel{\labelthis{rule:b}{b}}
\UnaryInfC{$\hvderiv{(\lambda x.\, x)}{1}{t}{T}$}
\DisplayProof

\\[4ex]
\AxiomC{$\hvderiv{f}{f'}{t}{T}$}
\AxiomC{$\hvderiv{g}{g'}{t}{T}$}
\RightLabel{\labelthis{rule:c}{c}}
\BinaryInfC{$\hvderiv{(\lambda x. (f~x, g~x))}{(f', g')}{t}{T}$}
\DisplayProof
\qquad
\AxiomC{$\hvderiv{f}{f'}{t}{T}$}
\AxiomC{$\hvderiv{g}{g'}{t}{T}$}
\RightLabel{\labelthis{rule:d}{d}}
\BinaryInfC{$\hvderiv{(\lambda x. f~x + g~x)}{f'+ g'}{t}{T}$}
\DisplayProof
\\[4ex]
\AxiomC{$\hvderiv{f}{f'}{t}{T}$}
\RightLabel{\labelthis{rule:e}{e}}
\UnaryInfC{$\hvderiv{(\lambda x. \sin(f~x))}{(f' \cdot \cos(f~t)))}{t}{T}$}
\DisplayProof
\\[4ex]
\AxiomC{$\hvderiv{f}{f'}{t}{T}$}
\AxiomC{$f~t > 0$}
\RightLabel{\labelthis{rule:f}{f}}
\BinaryInfC{$\hvderiv{(\lambda x. \sqrt{f~x})}{(f' \cdot 1 / (2 \cdot \sqrt{f~t}))}{t}{T}$}
\DisplayProof
\\[4ex]
\AxiomC{$\hvderiv{f}{f'}{t}{T}$}
\AxiomC{$\hvderiv{g}{g'}{t}{T}$}
\AxiomC{$g~t \neq 0$}
\RightLabel{\labelthis{rule:g}{g}}
\TrinaryInfC{$\hvderiv{(\lambda x. f~x / g~x)}{(- f~t \cdot (1 / (g~t) \cdot g' \cdot 1 / (g~t)) + f' / g~t))}{t}{T}$}
\DisplayProof
\end{tabular}
\end{theorem}
These are standard laws, but in a deductive rather than equational form. A constant function $\lambda x.\, c$ has derivative $0$~\eqref{rule:a}, and the identity function $\lambda x. \, x$ has derivative $1$~\eqref{rule:b}. If a derivative is a composed of a pair $(f', g')$ then it can be decomposed into two derivative proofs~\eqref{rule:c}. This law is particularly useful for decomposing a SODE into its component ODEs. A function composed of two summed components can similarly be composed~\eqref{rule:d}. The derivative of $\sin$ is $\cos$~\eqref{rule:e}. The remaining two rules are for square root~\eqref{rule:f} and division~\eqref{rule:g}. They both have additional provisos to avoid undefinedness. Square root $\sqrt{x}$ can be differentiated only when $x > 0$. Similarly, a division requires that the denominator is non-zero, hence the extra proviso.

The strategy employed by \texttt{ode\_cert} is as follows:

\begin{nonformat-algorithm}[SODE Certification Method] $ $%

\begin{enumerate}
    \item Decompose a SODE in $n$ variables to $n$ subgoals of the form $\hvderiv{f_i}{f_i'}{t}{T}$ for $1 \le i \le n$ \label{it:step1};
    \item Replace every such goal with two goals: $\hvderiv{f_i}{X_i}{t}{T}$ and $X_i = f_i'$ using a fresh meta-variable $X_i$. The latter goal is used to prove equivalence between the expected and actual derivative in $f'$;
    \item For each remaining derivative goal, recursively apply the derivative introduction laws (Theorem~\ref{thm:der-intro}). If any derivative goals remain, the method fails;
    \item The remaining subgoals are equalities and inequalities in the real variables. Attempt to discharge them all using real arithmetic and field laws using the simplifier tactic for recursive equational rewriting.
    \item If no goals remain, the ODE is certified.
\end{enumerate}
\end{nonformat-algorithm}
We exemplify this method with Example~\ref{ex:SODE}, using $f \defs (\lambda t~(x, y, z). (t, x, 1)$. A proposed solution is $x \defs (\lambda t. (t^2 / 2 + x_0, t^3 / 6 + x_0 \cdot t + y_0, z_0 + t))$, where $x_0, y_0, z_0$ are integration constants, or initial values for variables. We form the goal $x \mathop{\skey{solves-ode}} f~\real~\real^3$ and execute \texttt{ode\_cert}. The domain and range constraints are trivial in this case. Following step~\eqref{it:step1}, we obtain 3 subgoals: 

\begin{enumerate}
    \item $\hvderiv{(\lambda t.\, t^2 / 2 + x_0)}{t}{t}{T}$;
    \item $\hvderiv{(\lambda t.\, t^3 / 6 + x_0 \cdot t + y_0)}{x}{t}{T}$;
    \item $\hvderiv{(\lambda t.\, z_0 + t)}{1}{t}{T}$
\end{enumerate}
We focus on the second subgoal. Having applied the derivative introduction laws, we receive two proof obligations. The first is $6 \neq 0$, which is required since $6$ is the denominator in a division, and is trivial. The second is the following equality:
$$- (t^3) \cdot (1/6 \cdot 0 \cdot 1/6) + 3 \cdot 1 \cdot t^{3-1} / 6 + (x_0 \cdot 1 + 0 \cdot t) + 0 = t^2 / 2 + x_0$$
Though seemingly complex, it simplifies to give the desired result, since all but one of the summands reduce to $0$. This, and more complex goals, can be solved using the built-in simplification sets \texttt{algebra_simps} and \texttt{field_simps}. The other two derivative subgoals similarly reduce, and so the solution is certified.

The interface for the CAS tools is through two Isabelle commands:

\begin{alltt}
\textbf{ode_solve} <SODE>
\textbf{ode_solve_thm} (<NAME>:)? <SODE> <DOM>? <CODOM>? <ASSM>?
\end{alltt}

\noindent The \texttt{\textbf{ode_solve}} command takes a SODE in the form used in Example~\ref{ex:SODE}, and sends this to the CAS for processing. If a solution is found, the tool suggests a lemma that can be inserted of the form $x \mathop{\skey{solves-ode}} f~T~D$, with a concrete solution $x$, in the style of the \textsf{sledgehammer} tool~\cite{Blanchette2011,Blanchette2016Hammers}. The given lemma is proved using \texttt{ode\_cert}. \texttt{\textbf{ode_solve_thm}} produces a lemma directly, with the given name. It also optionally allows specification of an explicit domain, codomain, and assumption. The assumption is necessary if the SODE contains constants that are locally constrained. Different CAS systems can be selected using the Isabelle variable \texttt{SODE_solver}, which can take the value \texttt{fricas}, \texttt{maxima}, \texttt{sympy}, or \texttt{wolfram}. An example showing our tool can be seen in Figure~\ref{fig:isabelle-screenshot}.

In the next two sections we describe our integrations of Isabelle with SageMath and the Wolfram Engine.

\section{SageMath Integration}
\label{sec:sagemath}

\begin{wrapfigure}{r}{0.4\linewidth}
    \vspace{-10ex}
    \centering
    \includegraphics[width=4.5cm]{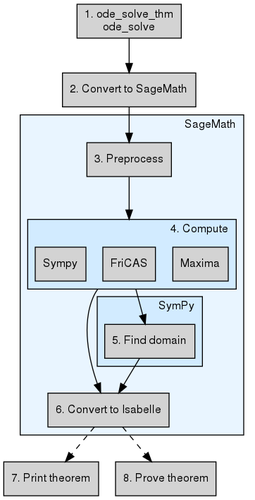}

    \vspace{-1ex}
    
    \caption{Overview of the SageMath pipeline}
    \label{fig:sageoverview}

    \vspace{-5ex}
\end{wrapfigure}

SageMath~\cite{sagemath} is an open source competitor to the Wolfram Engine. Its functionality is accessed via calls to a Python API. It integrates several open source CAS systems in order to provide its functionality, in each case choosing the best implementation for a particular symbolic computation. This makes SageMath an ideal target for integration with Isabelle. In the latest version of SageMath (version 9.1), Maxima~\cite{maxima} is the default CAS for solving SODEs. FriCAS~\cite{FriCAS} is also an option, though this is not bundled with SageMath by default. Our plugin also supports the CAS SymPy~\cite{sympy}, which is implemented using the SymPy to SageMath translation functions.

An overview of the SageMath pipeline is shown in Figure~\ref{fig:sageoverview}. The distinct steps that the SageMath integration uses are Steps 2 to 6.

\vspace{1ex}
\noindent \textbf{Step 2 and 6: Conversion.} In Step 2, the SageMath integration code receives a term for the input SODE. This is traversed, converted to a string containing Python code, passed to a Python integration script over the command line, and evaluated using Python's \verb|eval| function.
In Step 6 the opposite happens: the Python integration script traverses the expression, converts it to a string containing Isabelle code, and returns it on standard output. This string is then evaluated using the Isabelle function \verb|Syntax.read_prop : context -> string -> term|, which parses and type checks a proposition term in a given proof context.

Converting between the two representations is mostly a task of mapping between function names. However, there are several exceptions to this rule:

\begin{enumerate}
    \item Numbers in Isabelle are in a decomposed binary format. The Isabelle function \verb|HOLogic.dest_number| is used to convert these to integer values.
    \item There are different operators in Isabelle for integer, rational and real powers, whereas
    SageMath uses one operator. When converting from SageMath to Isabelle, the plugin chooses the simplest type of power function.
    \item Isabelle does not contain a representation of the mathematical constant $e$, but rather
    the exponential function $e^n$. When $e$ is used on its own in
    SageMath, this is converted into $\exp(1)$.
\end{enumerate}

\vspace{1ex}
\noindent \textbf{Step 3: Preprocessing.}
In many CAS systems, the SODE solving functionality is less powerful than the single equation ODE solving functionality. This often means that SODEs can be solved by the CAS system only when rewritten as ODEs. The preprocessing step, described in Algorithm~\ref{alg:preprocessing},  takes advantage of this by rewriting two different types of SODEs:
\begin{enumerate}
    \item SODEs formed from a higher order ODE, where a variable is introduced to represent a higher derivative, so that the SODE can conform to the format specified in \verb|ode_cert| (for example $(\dot{x},\dot{y}) = (2x+y, x)$). This can be preprocessed back into the higher order ODE which was originally intended.
    \item SODEs formed from two distinct system. For example, the SODE of a particle acting under gravity with a constant horizontal velocity - $(\dot{v}_x,\dot{v}_y) = (2, -g)$. This can be preprocessed into multiple independent ODEs, and solved using the CAS's ODE solving functionality.
\end{enumerate}

\setcounter{algocf}{1}
\begin{algorithm}[t]
\SetKwData{Sode}{sode}
\SetKwData{Equation}{equation}
\SetKwData{SimpleEquations}{simple-equations}
\SetKwFunction{SolveODE}{SolveODE}
\SetKwFunction{SolveSODE}{SolveSODE}
\SetKwData{SolvedEquation}{solved-equation}
Initialise \SimpleEquations to an empty array\;
\Repeat{\Sode is unchanged}{
    \ForEach{\Equation in the \Sode}{
      \If{\Equation is of the form $\dot{y}=x$ for any variable $x$ or $y$}{
            replace any occurrence of $x$ in \Sode with $\dot{y}$\;
            append \Equation to \SimpleEquations\;
            remove \Equation from \Sode
        }
      \If{\Equation is of the form $\dot{y}=f(t, x)$ (the equation is solely a function of it's independent and dependent variables)}{
            \SolvedEquation $\leftarrow$ \SolveODE{\Equation}\;
            replace any occurrence of $y$ in \Sode with \SolvedEquation\;
            remove \Equation from \Sode\;
            output \SolvedEquation as a solution to \Equation
        }
    }
}
\BlankLine
solve \Sode and output the solutions\;
\BlankLine
\ForEach{\Equation as $\dot{y}=x$ in \SimpleEquations}{
    find the solution to $\dot{x}$ from the existing outputs, differentiate it and output this as a solution to $\dot{y}$
}

\caption{Preprocessing Step Algorithm}
\label{alg:preprocessing}
\end{algorithm}

\noindent As an example of Algorithm~\ref{alg:preprocessing}, consider again Example~\ref{ex:SODE}
($(\dot{x}, \dot{y}, \dot{z}) = (t, x, 1)$). We can apply the first rewriting rule, as the equation $\dot{y} = x$ exists
in this SODE. This means we can transform the equation $\dot{x} = t$ into $\ddot{y} = t$, yielding a new SODE of the
form $(\ddot{y}, \dot{z}) = (t, 1)$. The two equations ($\ddot{y} = t$ and $\dot{x} = 1$) in this SODE are expressed
solely in terms of their independent and dependent variables, so they can be solved without considering the other
equations. This yields the solution:

\begin{equation}
(y, z) = \left(\frac{t^3}{6} + c_0t + c_1, t + c_2\right)
\end{equation}

\noindent We can now find $x$ by calculating $\dot{y}$. The final solution is:

\begin{equation}
(x, y, z) = \left(\frac{t^2}{3} + c_0, \frac{t^3}{6} + c_0t + c_1, t + c_2\right)
\end{equation}

\vspace{1ex}
\noindent\textbf{Step 4: Solving.} In Step 4, the input SODE is fed into one of three SODE solvers: SymPy, Maxima and FriCAS. These three solvers were all considered as potential CAS systems to use, in the order of: SymPy then Maxima then FriCAS. FriCAS performs best on the test set, but the option to use the other CAS systems is preserved.

\vspace{1ex}
\noindent\textbf{Step 5: Domain finding.} When verifying the solution of a SODE, \verb|ode_cert| requires a domain for which the solution is valid. When SageMath returns a solution, it does not return this information, therefore this domain needs to be generated from the solution. The strategy we have taken is to assume the domain for which the solution is valid to be the maximal domain of the solution.

SageMath does not have any maximal domain finding functionality, so we have used SymPy for this part of the pipeline. The function that finds maximal domains in SymPy was also patched to ensure that greater maximal domains can be calculated\footnote{The pull request for this can be found at \url{https://github.com/sympy/sympy/pull/19024}. This is merged at \url{https://github.com/sympy/sympy/pull/19047}}.

We evaluate our SageMath integration in \S\ref{sec:evaluation}, but first, in the next section, we describe our Wolfram integration.

\section{Wolfram Engine Integration}
\label{sec:wolfram}

\begin{figure}
\begin{center}
\begin{tikzpicture}[->,node distance=0.8cm, auto]
     \node[supp] (isa) {ODE term};
     \node[supp, above=of isa, yshift=2cm] (mat) {Wolfram expression};
     \node[supp, right=of mat, xshift=1.5cm] (matS) {Wolfram solution};
     \node[supp, below=of matS](ast){Expression datatype};
     \node[supp, right=of isa] (sol) {Isabelle solution};
     \node[supp, below=of ast, right=of sol] (dom) {Solution domain};
     \node[main, below=of sol] (lemma) {Solution lemma};
     \draw[arrow]  ([xshift=-1cm]isa.west) node[above]{\texttt{\textbf{ode_solve}}} -- (isa.west);
     \draw[arrow]  (lemma.south) node[right, yshift=-0.5cm]{Lemma} -- ([yshift=-1cm]lemma.south);
     \path[arrow]
        (isa) edge node[left]{Translation}(mat)
        (mat.north) edge[bend left] node [above]{Wolfram Engine interface} (matS.north)
        (matS) edge node{Lex and parse} (ast)
        (ast) edge node[right, align=center]{Wolfram Engine\\interface} (dom)
        (ast) edge node[left]{Interpret}(sol)
        (isa) edge node[left, yshift=-0.4cm]{Unparse}(lemma)
        (sol) edge (lemma)
        (dom) edge node[right, yshift=-0.4cm, text width=3cm]{Lex and parse,\\interpret} (lemma);
\end{tikzpicture} 
\caption{Wolfram plugin workflow}\label{fig:workflow}
\end{center}
\end{figure}
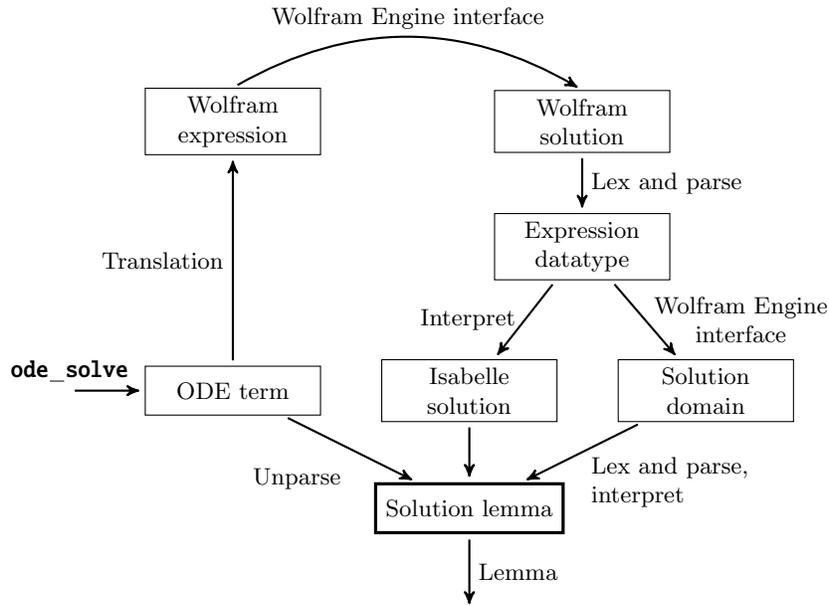

Using the Wolfram Engine over SageMath comes with the main disadvantage of
losing open-source status, but it also has a few advantages.
In our implementation, it is notably faster at producing solutions, and the
SODEs require no preprocessing before solving. Our Wolfram interface is written
entirely in SML, which makes it easier for those familiar with Isabelle system
programming to use and extend.
 
The implementation of the Wolfram interface is illustrated in Figure~\ref{fig:workflow}.
First, the Isabelle term that represents the SODE is translated to an equivalent Wolfram expression. This is passed to the Wolfram
Engine for solving. The Wolfram solution to the SODE is lexed and parsed, and stored as an ML
datatype. The Wolfram interface is used to retrieve and parse the solution
domain, and all parsed expressions are translated to Isabelle. Finally, the
plugin combines the domain, the solution, and the original SODE to produce the
solution theorem.

Default Wolfram expressions are typeset and difficult to parse, so we instead
retrieve solutions from the Wolfram Engine in ``full form''\footnote{Please see \url{https://reference.wolfram.com/language/ref/FullForm.html}.}. This format presents the expression in a similar style to an algebraic datatype, with explicit constructors, and is implemented in our tool as the following ML datatype.

\begin{alltt}
\textbf{datatype} expr = Int \textbf{of} int | Real \textbf{of} real | Id  \textbf{of} string |
  Fun \textbf{of} string * expr list | CurryFun \textbf{of} string * expr list list
\end{alltt}
\noindent We distinguish between functions with only one set of arguments (\texttt{Fun}) and those
with several (\texttt{CurryFun}), as the latter are uncommon and dealing with them
clutters the code.

To illustrate the implementation stages, the internal representation at each
stage is shown for Example~\ref{ex:SODE}.
The approach for translation of the SODE to Wolfram is the following:
\begin{enumerate}
    \item Generate an alphabetically ordered variable mapping, for each of the SODE variables, to avoid name clashes and ease solution reconstruction.
    \item Translate the term to an equivalent Wolfram expression by traversing the expression tree.
    \item Construct a \texttt{DSolve} call using the expression.
\end{enumerate}

\noindent \texttt{DSolve} is the general differential equation solver for the Wolfram Engine~\cite{wolfram}, which can solve a list of differential equations for given dependent and independent variables.

We exemplify the translation, again using Example~\ref{ex:SODE}. To represent this system, the following variable mapping is used:\[t \rightarrow a,
x\rightarrow b, y\rightarrow c, z \rightarrow d\]

\noindent Using this mapping, the system is translated to the following \texttt{DSolve} call:

\begin{alltt}
DSolve[
    \{b'[a]==a, c'[a]==b[a], d'[a]==1\},
    \{b[a],c[a],d[a]\},
    a]
\end{alltt} 

\noindent The Wolfram engine is called using its command-line interface \texttt{wolframscript}, which takes a function call as an argument and returns the result. Warnings are suppressed to facilitate parsing.
The Wolfram engine represents solutions as a list of rules, which are simply functions on expressions. Many solutions may be returned, but we only use the first one, which is lexed and parsed. The maximal domain of this solution is retrieved in another call to the Wolfram Engine, similar to the SageMath integration.

The solution to the test ODE after lexing and parsing is the following:

\begin{alltt}
Fun ("List",
    [Fun ("List",
          [Fun ("Rule",
                [Fun ("b", [Id "a"]), Fun ("Plus", ...)]),
           Fun ("Rule",
                [Fun ("c", [Id "a"]), Fun ("Plus", ...)]),
           Fun ("Rule",
                [Fun ("d", [Id "a"]), Fun ("Plus", ...)])])])
\end{alltt}
Here, the inner most list gives values for each of the continuous variables. The translation from such a Wolfram expression to an Isabelle term is done by reversing the variable mapping and then traversing the expression tree. There are special cases for the constant $e$, which is translated to the exponent function, and negative powers, similar to the SageMath integration. Solutions may be provided by the Wolfram Engine which use functions not available in Isabelle, such as those containing integrals. These are reported as errors. Finally, the lemma is assembled by combining the domain, solution, and original SODE.

This completes our description of the two CAS integrations. In the next section we evaluate them both.

\section{Evaluation}
\label{sec:evaluation}

In this section, we evaluate our approach to certifying SODEs. We consider two test sets, to which we apply both CAS integrations, and then evaluate the results.

The first test set is generated programmatically by searching the KeYmaera X example repository\footnote{These examples can be found here: \url{https://github.com/LS-Lab/KeYmaera-release/tree/master/examples/hybrid}} for any lines containing
fragments of the form \verb|{<EQUATION>}|, which describes of a SODE in KeYmaera X. Any duplicate SODEs are then combined. In total, 148 SODEs were found, 79 were duplicate pairs, leaving 69 unique SODEs to add to the test set. To represent these equations, and from now on, $t$ is the independent variable, meaning $(\dot{x}, \dot{y}) = \left(\frac{dx}{dt},\frac{dy}{dt}\right)$: The KeYmaera X tests contains many ``simple" SODEs. For example, the basic gravity SODE:
\begin{equation}
(\dot{h}, \dot{v}) = (v, -g)
\end{equation}
\noindent was present in many variations. A few of the test cases also contains more complex SODEs, such as the following example taken from the test set:
\begin{equation}
(\dot{q_x}, \dot{q_y}, \dot{f_x}, \dot{f_y}) = \left(f_x\frac{Kq_x}{D}, f_y\frac{Kq_y}{D}, f_{xp},
f_{yp}\right) \label{eq:KYXSODE}
\end{equation}

We class complex SODEs as those that contain at least one operator, excluding unary minus (for example $-2$). For example,  Equation~\ref{eq:KYXSODE} contains four: two multiplication operators and two division operators. Under this classification, this test set contains 20 ``complex" SODEs and 49 ``simple" SODEs.

\begin{figure}[t]
    \centering
    \includegraphics[width=.8\linewidth]{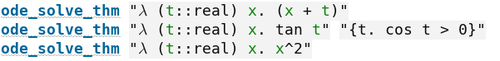}
    \vspace{-2ex}
    
    \caption{Wolfram SODE Solver Integration in Isabelle/HOL}
    \label{fig:isabelle-screenshot}
    
    \vspace{-4ex}
\end{figure}

The KeYmaera X test set is restricted by the capabilities of KeYmaera X, which doesn't support transcendental functions, and also its typical applications -- cyber-physical systems. In order to test the full capabilities of our tool, we construct a second test set. For this, we try to cover a wide variety of cases, to highlight any areas where our tool might have problems. Table~\ref{tab:testCases} contains this test set, together with the rationale behind why each expression was chosen. Figure~\ref{fig:isabelle-screenshot} shows some of these test cases running in the tool.

\begin{table}[t]
\caption{Test cases}

\vspace{-4ex}

\begin{center}
\begin{tabular}{ |c| >{\centering\arraybackslash}m{10em} |c| }
    \hline
    Number & ODE System: \(\dot{x}, (\dot{y}, \dots) = \dots\) & Rationale \\ \hline
    1  & $x+t$ & Inhomogeneous polynomial\\ 
    2  & $\tan(t)$ & Tangent function \\ 
    3  & \(x^2\) & Second order polynomial\\
    4  & \(-y, x\) & Trigonometric solution\\
    5  & $1/t$ & Domain issues at 0\\ 
    6  & $1/(2x - 1)$ & Has two solutions \\ %
    7  & $xy, 3$ & Contains a factor of $xy$ \\ 
    8  & $2x+y, x$ & Homogeneous \nth{2} order SODE \\ 
    9 & $2x+y+t^2, x$ & Inhomogeneous \nth{2} order SODE \\ 
    10 & $\arcsin(t)$ & Inverse trigonometric function \\ 
    11 & \(\sqrt{t}\) & Square root\\
    12 & $\sqrt[5]{t}$ & Higher roots \\ 
    13 & $t^{\sqrt{2}}$ & Non-rational powers \\ 
    14 & $x + y, y + 2z, x^2+1$ & Higher dimensional SODE \\
    15 & \(x^2 - t\) & Bessel function\\
    16 & \(y, e^{t^2}\) & Imaginary error function\\
    17 & \(\sin(x)/\ln(x)\) & Impossible to solve\\
    18 & \(\ln(t), x\) & Logarithmic\\\hline
\end{tabular}
\label{tab:testCases}
\end{center}

\vspace{-5ex}

\end{table}

Most of the KeYmaera test cases were successfully solved by both SageMath/FriCAS+ and Wolfram using \verb|ode_cert|, apart from five test cases:
\begin{align}
    \dot{x} &= x^2+x^4\\
    (\dot{x}, \dot{y}) &= ({x-3}^4+y^5, y^2)\\
    \dot{x} &= {x-3}^4+a\\
    (\dot{x}, \dot{y}) &= (v, g+d v^2) \label{eqn:KYMFail4} \\
    (\dot{x}, \dot{y}) &= (y, -w^2 x-2 d w y) \label{eqn:KYMFail5}
\end{align}
\noindent Both SageMath/FriCAS+ and Wolfram could not solve the first three equations. In KeYmaera X~\cite{KeYmaera} and also Isabelle/HOL~\cite{Foster2020-dL}, such SODEs can be verified using differential induction~\cite{dL} rather than explicit solutions. Equation~\ref{eqn:KYMFail4} was solved correctly only by Wolfram Engine, but \verb|ode_cert| was unable to prove this and Equation~\ref{eqn:KYMFail5} was solved by both Wolfram and SageMath, but \verb|ode_cert| was again unable to certify it. In all the cases where \verb|ode_cert| was unable to prove the the result, this was due to a failure to prove a large algebraic proposition containing more than 50 operator applications. %

In addition, 7 of the test cases required an assumption to be specified in the \verb|ode_solve_thm| statement. For example, the SODE $$(\dot{x}, \dot{t}) = (c+b(u-x), 1)$$ requires the assumption $b > 0$. This means that out of the 20 ``complex" test cases, 15 could be automatically solved and verified and all of the 49 ``simple" test cases could be automatically solved and verified.

\begin{table}[t]
    \caption{Results from the SODE tests.}
    
    \vspace{-4ex}
    
    \begin{center}
        \begin{threeparttable}
        \begin{tabular}{
            |c|>{\centering\arraybackslash}m{50pt}
            >{\centering\arraybackslash}m{50pt}
            >{\centering\arraybackslash}m{50pt}
            |>{\centering\arraybackslash}m{50pt}
            >{\centering\arraybackslash}m{50pt}
            >{\centering\arraybackslash}m{50pt}|}
            \hline
        Number & Solved by the CAS & Correct domain found
            & Proved automatically in Isabelle &
            Solved by the CAS & Correct domain found
            & Proved automatically in Isabelle\\
            \hline
            & \multicolumn{3}{c|}{SageMath \footnotemark[1]}
            & \multicolumn{3}{c|}{Wolfram}\\
        \hline
1       & \yes & \yes & \yes     & \yes & \yes & \yes \\
2       & \yes & \no & \no       & \yes & \no & \yes  \\
3       & \yes & \yes & \yes     & \yes & \yes & \yes \\
4       & \yes & \yes & \yes     & \yes & \yes & \yes \\
5       & \yes & \yes & \yes     & \yes & \yes & \yes  \\
6       & \yes & \no & \no       & \yes & \no & \no \\
7       & \yes & \yes & \yes     & \yes & \yes & \yes \\
8       & \yes & \no & \no       & \yes & \no & \no \\
9       & \yes & \no & \no       & \no & \no & \no \\
10      & \yes & \no & \no       & \yes & \no & \no \\
11      & \yes & \no & \yes      & \yes & \no & \yes \\
12      & \yes & \no & \yes      & \yes & \no & \yes \\
13      & \yes & \no & \no       & \yes & \no & \no \\
14      & \no  & N/A  & N/A      & \no  & N/A  & N/A \\
15      & \no  & N/A  & N/A      & \no  & N/A  & N/A \\
16      & \no  & N/A  & N/A      & \no  & N/A  & N/A  \\
17      & \no  & N/A  & N/A      & \no  & N/A  & N/A  \\
18      & \yes & \yes & \yes     & \yes & \yes & \yes  \\

\hline
        \end{tabular}
        \begin{tablenotes}[flushleft]
            \small 
            \item[1] This refers to using SageMath/FriCAS with the preprocessing step.
        \end{tablenotes}
    \end{threeparttable}
    \end{center}
    \label{table:ourSODEtestResults}
    
    \vspace{-5ex}
\end{table}

The results from our additional SODE test cases are presented in Table~\ref{table:ourSODEtestResults}. In these results, 10 of the 18 test cases could not be automatically proved by Isabelle. There are four distinct reasons behind these failures:
\begin{enumerate}
    \item The tactic \verb|ode_cert| cannot automatically prove the stated theorem. In all of the test cases where this occurs, this is due to \verb|ode_cert| failing to prove an algebraic proposition. This occurs in test case 2 for SageMath's result; and 8, 9, 10, 13 for both CAS systems. We have been able to prove test cases 8, 9 and 2 correct manually with the help of the \textsf{sledgehammer} tool\footnote{This work can be found here: \url{https://github.com/ThomasHickman/Isabelle-CAS-Integration/blob/master/manually_solved_cases.thy}}~\cite{Blanchette2016Hammers}. The goals left to prove in cases 8, 9, 10 and 13 contained more than 50 operators.
    \item The CAS system cannot produce the correct answer, but if it did, Isabelle could not prove the answer, as appropriate derivative laws have not been implemented. This occurs in test cases 15 and 16.
    \item The CAS system cannot produce the correct answer, and it is theoretically impossible for it to do so. This occurs in test case 17.
    \item The CAS system cannot produce the correct answer, but Isabelle does contain the derivative laws to prove the answer, if one was produced. This occurs in test case 14.
\end{enumerate}

\noindent Consequently, if we exclude the cases where the CAS system cannot provide a solution, and include those where a proof using \textsf{sledgehammer} was required, the rate of success is 11 out of 14, with 3 uncertifiable solutions.
  
\section{Conclusions}
\label{sec:concl}

In this paper, we described our work on integrating Isabelle with SODE solving in SageMath and Wolfram, to support verification of hybrid systems. In \S\ref{sec:ode-cert} we introduced the tactic \verb|ode_cert| for the automatic certification of SODE solutions. In \S\ref{sec:sagemath} and \S\ref{sec:wolfram} we described our integration with SageMath and Wolfram, respectively. In \S\ref{sec:evaluation} we evaluated our plugin using two different test sets: one generated from the KeYmaera X examples, and one with more complex examples.

We consider our approach to be successful. Our integration managed to solve and prove most of the test cases that we have devised. This means that projects using SODEs with a similar scope to those in the KeYmaera X examples should be able to use our plugin to generate certified solutions. This largely down to the impressive library of theorems developed in \textsf{HOL-Analysis}~\cite{Harrison2005-Euclidean} and \textsf{HOL-ODE}~\cite{immler2012,immler2018}, which allow certification to be substantially automated.

However, as we have noted five of our test cases (two from KeYmaera X, and three of our own) produced solutions that
could not be certified. This could either be due to lack of proof rules for derivation and real arithmetic in
Isabelle. Alternatively, it could be that the solutions returned by the CAS systems are in reality approximations, as
indicated by their size compared to the actual SODE. We plan to investigate this further in the future. Either way, we
note that Isabelle places a high bar on the artifacts that are accepted as mathematically sound, which gives confidence that
they can be used in safety critical applications.

In future work, we plan to use our integration as part of a Isabelle-based hybrid systems verification tool, using our implementation of \dL and related calculi~\cite{Foster2020-dL,Foster2020-Marine}. We aim to apply to a number of example projects, such as the KeYmaera X examples. This may expose areas in the Isabelle/HOL hybrid systems infrastructure that require improvement. In addition, we will investigate the integration of other CAS features into Isabelle. One example of this is quantifier elimination, which would further improve automation.

\vspace{1ex}

\noindent \textbf{Acknowledgements.} This work is supported by the EPSRC-UKRI Fellowship project \textit{CyPhyAssure}, grant reference EP/S001190/1.

\bibliographystyle{splncs}
\bibliography{references}

\begin{thebibliography}{10}

\bibitem{alur2011}
Alur, R.:
\newblock Formal verification of hybrid systems.
\newblock In: Proc. 9th. ACM Intl. Conf. on Embedded Software (EMSOFT), New
  York, NY, USA, ACM (2011)  273--278

\bibitem{KeYmaera}
Fulton, N., Mitsch, S., Quesel, J.D., V{\"o}lp, M., Platzer, A.:
\newblock {KeYmaera X}: An axiomatic tactical theorem prover for hybrid
  systems.
\newblock In Felty, A.P., Middeldorp, A., eds.: CADE. Volume 9195 of LNCS.,
  Springer (2015)  527--538

\bibitem{Gleirscher2018-NewOpportunitiesIntegrated}
Gleirscher, M., Foster, S., Woodcock, J.:
\newblock New opportunities for integrated formal methods.
\newblock ACM Comput. Surv. \textbf{52}(6) (2019)

\bibitem{Isabelle}
Nipkow, T., Wenzel, M., Paulson, L.C.:
\newblock {Isabelle/HOL: A Proof Assistant for Higher-Order Logic}. Volume 2283
  of LNCS.
\newblock Springer (2002)

\bibitem{Wenzel2007FMIsabelle}
Wenzel, M., Wolff, B.:
\newblock Building formal method tools in the {Isabelle/Isar} framework.
\newblock In: TPHOLs. Volume 4732 of LNCS., Springer (2007)

\bibitem{Brucker2019-DOFCert}
Brucker, A., Wolff, B.:
\newblock Using ontologies in formal developments targeting certification.
\newblock In: iFM. Volume 11918 of LNCS., Springer (2019)  65--82

\bibitem{Foster2020AUV}
Foster, S., Nemouchi, Y., O'Halloran, C., Tudor, N., Stephenson, K.:
\newblock Formal model-based assurance cases in {Isabelle/SACM}: An autonomous
  underwater vehicle case study.
\newblock In: FormaliSE, ACM (2020)

\bibitem{Harrison2005-Euclidean}
Harrison, J.:
\newblock A {HOL} theory of {E}uclidean space.
\newblock In Hurd, J., Melham, T., eds.: Theorem Proving in Higher Order
  Logics, 18th International Conference, TPHOLs 2005. Volume 3603 of LNCS.,
  Oxford, UK, Springer (August 2005)

\bibitem{immler2018}
Immler, F.:
\newblock A verified {ODE} solver and the {L}orenz attractor.
\newblock J. Autom. Reasoning \textbf{61}(1)  73--111

\bibitem{immler2012}
Fabian, I., H{\"o}lzl, J.:
\newblock Numerical analysis of ordinary differential equations in
  {I}sabelle/{HOL}.
\newblock In Beringer, L., Felty, A., eds.: ITP. Volume 7406 of LNCS., Springer
  (2012)  377--392

\bibitem{Blanchette2011}
Blanchette, J.C., Bulwahn, L., Nipkow, T.:
\newblock Automatic proof and disproof in {Isabelle/HOL}.
\newblock In: FroCoS. Volume 6989 of LNCS., Springer (2011)  12--27

\bibitem{Blanchette2016Hammers}
Blanchette, J.C., Kaliszyk, C., Paulson, L.C., Urban, J.:
\newblock Hammering towards {QED}.
\newblock Journal of Formalized Reasoning \textbf{9}(1) (2016)

\bibitem{dL}
Platzer, A.:
\newblock Differential dynamic logic for hybrid systems.
\newblock J. Autom. Reas. \textbf{41}(2) (2008)  143--189

\bibitem{morerobix}
Mitsch, S., Ghorbal, K., Vogelbacher, D., Platzer, A.:
\newblock Formal verification of obstacle avoidance and navigation of ground
  robots.
\newblock The International Journal of Robotics Research \textbf{36}(12) (2017)
   1312--1340

\bibitem{acasx}
Jeannin, J.B., Ghorbal, K., Kouskoulas, Y., Schmidt, A., Gardner, R., Mitsch,
  S., Platzer, A.:
\newblock A formally verified hybrid system for safe advisories in the
  next-generation airborne collision avoidance system.
\newblock Software Tools for Technology Transfer \textbf{19}(6)  717--741

\bibitem{Tuong2019-CIsabelle}
Tuong, F., Wolff, B.:
\newblock Deeply integrating {C11} code support into {Isabelle/PIDE}.
\newblock In: F-IDE. Volume 310 of EPTCS. (2019)  13--28

\bibitem{veriphy}
Bohrer, B., Tan, Y.K., Mitsch, S., Myreen, M.O., Platzer, A.:
\newblock {V}eri{P}hy: Verified controller executables from verified
  cyber-physical system models.
\newblock SIGPLAN Not. \textbf{53}(4) (2018)  617--630

\bibitem{modelplex}
Mitsch, S., Platzer, A.:
\newblock {ModelPlex}: Verified runtime validation of verified cyber-physical
  system models.
\newblock Form. Methods Syst. Des. \textbf{49}(1) (2016)  33--74 Special issue
  of selected papers from RV'14.

\bibitem{verified-dL}
Bohrer, B., Rahli, V., Vukotic, I., Platzer, A.:
\newblock Formally verified differential dynamic logic.
\newblock In Bertot, Y., Vafeiadis, V., eds.: Proc 6th ACM SIGPLAN Conf. on
  Certified Programs and Proofs (CPP), {ACM} (2017)  208--221

\bibitem{Munive2018-DDL}
Munive, J.H., Struth, G.:
\newblock Verifying hybrid systems with modal {Kleene} algebra.
\newblock In: RAMICS. Volume 11194 of LNCS., Springer (2018)

\bibitem{Foster2020-dL}
Munive, J.H., Struth, G., Foster, S.:
\newblock Differential {Hoare} logics and refinement calculi for hybrid systems
  with {Isabelle/HOL}.
\newblock In: RAMiCS. Volume 12062 of LNCS., Springer (April 2020)

\bibitem{Foster2020-IsabelleUTP}
Foster, S., Baxter, J., Cavalcanti, A., Woodcock, J., Zeyda, F.:
\newblock Unifying semantic foundations for automated verification tools in
  {Isabelle/UTP}.
\newblock Science of Computer Programming \textbf{197} (October 2020)

\bibitem{Li2017-Poly}
Li, W., Passmore, G., Paulson, L.:
\newblock Deciding univariate polynomial problems using untrusted certificates
  in {Isabelle/HOL}.
\newblock J. Autom. Reasoning \textbf{62} (2019)  29--91

\bibitem{sagemath}
{The Sage Developers}:
\newblock {SageMath, the Sage Mathematics Software System (Version 9.0)}.
  (2020)

\bibitem{maxima}
Maxima:
\newblock Maxima, a computer algebra system. version 5.34.1 (2014) Available at
  {http://maxima.sourceforge.net/}.

\bibitem{FriCAS}
{FriCAS team}:
\newblock {FriCAS---an advanced computer algebra system} (2019) Available at
  \url{http://fricas.sf.net}.

\bibitem{sympy}
Meurer, A., Smith, C.P., Paprocki, M., \v{C}ert\'{i}k, O., Kirpichev, S.B.,
  Rocklin, M., Kumar, A., Ivanov, S., Moore, J.K., Singh, S., Rathnayake, T.,
  Vig, S., Granger, B.E., Muller, R.P., Bonazzi, F., Gupta, H., Vats, S.,
  Johansson, F., Pedregosa, F., Curry, M.J., Terrel, A.R., Rou\v{c}ka, v.,
  Saboo, A., Fernando, I., Kulal, S., Cimrman, R., Scopatz, A.:
\newblock {SymPy}: symbolic computing in {Python}.
\newblock PeerJ Computer Science \textbf{3} (January 2017)  e103

\bibitem{wolfram}
{Wolfram Research, Inc.}:
\newblock Wolfram language documentation Available at
  \url{https://reference.wolfram.com}.

\bibitem{Foster2020-Marine}
Foster, S., Gleirscher, M., Calinescu, R.:
\newblock Towards deductive verification of control algorithms for autonomous
  marine vehicles.
\newblock In: 25th Proc. Intl. Conf. on Engineering of Complex Computer Systems
  (ICECCS), IEEE (March 2021)

\end{thebibliography}
 
\end{document}